\documentclass[fleqn,usenatbib]{mnras}
\pdfoutput=1

\usepackage[pdftex]{graphicx,color}
\usepackage[T1]{fontenc}
\usepackage[utf8]{inputenc}
\usepackage{lmodern}

\definecolor{urlblue}{rgb}{0,0,0.9}
\definecolor{linkgreen}{rgb}{0,0.45,0}
\definecolor{linkorange}{rgb}{0.7,0.1,0.0}

\usepackage{amsmath, amssymb}
\usepackage{cuted}
\usepackage[english]{babel}
\usepackage{enumerate}
\usepackage[normalem]{ulem}
\usepackage{enumitem}
\usepackage{multirow}
\usepackage{booktabs}
\usepackage{textcase}
\usepackage{makecell}

\setlist[enumerate]{wide=0pt, widest=99,leftmargin=\parindent, labelsep=* } 

\definecolor{valecol}{rgb}{0,0.5, 1.}

\definecolor{ziadcol}{rgb}{1.,.5, 0.0}

\graphicspath{{./figs/}}

\AtBeginDocument{\hypersetup{
linkcolor=linkgreen,
citecolor=linkorange,
urlcolor=urlblue}}

\title{A void in the Hubble tension? The end of the line for the Hubble bubble}

\author[Camarena, Marra, Sakr \& Clarkson]{
David Camarena,$^{1}$ Valerio Marra,$^{2,3,4}$ Ziad Sakr,$^{5}$ and Chris Clarkson$^{6,7,8}$
\\
$^{1}$PPGCosmo, Universidade Federal do Espírito Santo, 29075-910, Vitória, ES, Brazil\\
$^{2}$Núcleo de Astrofísica e Cosmologia \& Departamento de Física, Universidade Federal do Espírito Santo, 29075-910, Vitória, ES, Brazil\\
$^{3}$INAF -- Osservatorio Astronomico di Trieste, via Tiepolo 11, 34131, Trieste, Italy\\
$^{4}$IFPU -- Institute for Fundamental Physics of the Universe, via Beirut 2, 34151, Trieste, Italy\\
$^{5}$Université St Joseph; Faculty of Sciences, Beirut, Lebanon\\
$^{6}$Department of Physics and Astronomy, Queen Mary University of London, UK\\
$^{7}$Department of Physics \& Astronomy, University of the Western Cape, Cape Town 7535, South Africa\\
$^{8}$Department of Mathematics \& Applied Mathematics, University of Cape Town, Cape Town 7701, South Africa
}

\begin{document}

\label{firstpage}
\pagerange{\pageref{firstpage}--\pageref{lastpage}}

\maketitle

\begin{abstract}
The Universe may feature large-scale inhomogeneities beyond the standard paradigm, implying that statistical homogeneity and isotropy may be reached only on much larger scales than the usually assumed $\sim$100 Mpc.
This means that we are not necessarily 
typical
observers and that the Copernican principle could be recovered only on 
super-Hubble scales.
Here, we do not assume the validity of the Copernican principle and let Cosmic Microwave Background, Baryon Acoustic Oscillations, type Ia supernovae, local $H_0$, cosmic chronometers, Compton y-distortion and kinetic Sunyaev–Zeldovich observations constrain the geometrical degrees of freedom of the local structure, which we parametrize via the $\Lambda$LTB model---basically a non-linear radial perturbation of a FLRW metric.
In order to quantify if a non-Copernican structure could explain away the Hubble tension, we pay careful attention to computing the Hubble constant in an inhomogeneous universe, and we adopt 
model selection via both the Bayes factor and the Akaike information criterion.
Our results show that, while the $\Lambda$LTB model can successfully explain away the $H_0$ tension, it is favored with respect to the $\Lambda$CDM model only if one solely considers supernovae in the redshift range that is used to fit the Hubble constant, that is, $0.023<z<0.15$.
If one considers all the supernova sample, then the $H_0$ tension is not solved and the support for the $\Lambda$LTB model vanishes. Combined with other data sets, this solution to the Hubble tension barely helps. 
Finally, we have reconstructed our local spacetime. We have found that data are best fit by a shallow  void with $\delta_L \approx -0.04$ and $r^{\mathrm{out}}_L \approx 300$ Mpc, which, interestingly, lies on the border of the 95\% credible region relative to the standard model expectation.
\end{abstract}

\begin{keywords}
large-scale structure of Universe -- cosmology: observations -- cosmological parameters -- cosmology: theory
\end{keywords}

\section{Introduction} \label{sec:intro}

Accurate cosmological and astrophysical observations have revealed a discrepancy between early- and late-time determinations of the Hubble constant. This discrepancy, with a significance of $5\sigma$ if one considers CMB observations~\citep{Aghanim:2018eyx} and the local cosmic distance ladder~\citep{Riess:2021jrx}, is the so-called Hubble tension.
In the absence of unknown systematic errors, this discrepancy could suggest the existence of physics beyond the standard paradigm of cosmology. This scenario has led cosmologists to propose and study new cosmological models, mainly, but not limited to, those that extend the $\Lambda$CDM model at early- or late-times \citep[see][for an up-to-date and extensive review]{Abdalla:2022yfr}.

Although many of the models proposed to solve the Hubble tension involve modifications to dark matter and dark energy or changes to the theory of gravity, geometrical degrees of freedom have also been  considered.
Indeed, within the standard model, the universe is expected to be homogeneous and isotropic only at scales $\gtrsim$100 Mpc so that we may need to take into account the local perturbed spacetime when analyzing observations at low redshifts.
This may be relevant for the Hubble tension as the $H_0$ measurement of \citet{Riess:2021jrx} is based on the luminosity-distance-redshift relation in the redshift range $0.023<z<0.15$.

At the linear level, an adiabatic perturbation in the density of our local spacetime causes a perturbation in the expansion rate given by \citep{Marra:2013rba}:
\begin{equation} \label{linH}
\frac{\delta H_0}{H_0} = - \frac{1}{3} f(\Omega_m) \frac{\delta \rho(t_0)}{\rho(t_0)} \,,
\end{equation}
where $f\simeq 0.5$ is the present-day growth rate for the concordance $\Lambda$CDM model.
One can then see how a local underdensity, $\delta \rho/\rho <0$, would cause a higher local expansion rate, $\delta H_0/H_0 >0$.
However, perturbations are smaller on larger scales and the typical contrast -- i.e., dictated by the amplitude of perturbations as constrained by CMB observations within the standard $\Lambda$CDM model -- quickly decreases so that the homogeneous FLRW limit is reached.
Theoretical computations \citep[see][and references therein]{Camarena:2018nbr} and numerical simulations \citep[see][and references therein]{Odderskov:2017ivg} suggest that this cosmic variance on $H_0$ causes a 0.5--1\% systematic uncertainty when analyzing observations in the redshift range $0.023<z<0.15$, falling short of explaining the 9\% difference between early- and late-times constraints.

This failure in explaining away the Hubble tension is due to the fact that we assumed the standard spectrum of perturbations which is based on a series of assumptions, such as the Copernican principle, the use of the FLRW metric and standard slow-roll inflation.
However, the Universe may feature large-scale inhomogeneities beyond the standard paradigm, that is, statistical homogeneity and isotropy may be reached only on much larger scales than the usually assumed 100 Mpc. In other words, we are not necessarily 
typical 
observers and  the Copernican principle could be recovered only on grander scales so that observations could depend on the position of the observer and the notion of an average FLRW observer would cease to be meaningful \citep[][]{Kolb:2009rp}.
This could tremendously modify our perception of the cosmos and motivates us to take a pragmatic approach and test if a local inhomogeneity of any size and depth could solve the $H_0$ tension.

There has been growing observational evidence that the local universe is underdense on scales of several hundred megaparsecs, as reported by \citet{Frith:2003tb,Keenan:2013mfa,Whitbourn:2013mwa,Hoscheit:2018nfl,Haslbauer:2020xaa,Bohringer:2019tyj,Wong:2021fvu}.
Furthermore, several anomalous signals in cosmological observables have been emerging since the establishment of the $\Lambda$CDM model as the standard model of cosmology more than two decades ago.
Besides the Hubble crisis, particularly relevant here are the CMB anomalies and the cosmic dipoles \citep[see][and references therein]{Perivolaropoulos:2021jda}.
These signals are at odds with the standard paradigm according to which the spacetime is well described by the homogeneous and isotropic FLRW metric on scales larger than $\approx$100 Mpc.

In \citet{Camarena:2021mjr}, we tested if the  Copernican Principle is valid, that is, if we are indeed `typical' 
FLRW observers.
Specifically, we have probed radial inhomogeneity around us by constraining the $\Lambda$LTB model with the latest available data from CMB, BAO, type Ia supernovae, local $H_0$, cosmic chronometers, Compton y-distortion, and kinetic Sunyaev–Zeldovich effect.
The $\Lambda$LTB model is basically the $\Lambda$CDM model with the addition of an arbitrary spherical inhomogeneity.
We found that inhomogeneity around us approximately follows the expectation of the standard model.

Here, we extend the results of \citet{Camarena:2021mjr} in order to reconstruct our local spacetime and test its implications for the Hubble tension.
Special attention is given to the method used to measure the local Hubble constant in an inhomogeneous Universe, 
Bayesian model comparison, and a generalization of the LTB profile in order to better reconstruct our cosmological neighborhood.

Similar analyses using the $\Lambda$LTB model were carried out by \citet{Tokutake:2017zqf, Hoscheit:2018nfl, Kenworthy:2019qwq, Lukovic:2019ryg, Ding:2019mmw, Cai:2020tpy, Castello:2021uad}.
\citet{Kenworthy:2019qwq} looked at the luminosity distance-redshift relation of 1295 SNe over a redshift range of $0.01 < z < 2.26$ and concluded that data is inconsistent at the 4--5$\sigma$ confidence level with a large local underdensity with $\delta<-0.2$ so that local $H_0$ measurements are not affected by the local structure.
\citet{Lukovic:2019ryg} confronted luminosity data from 35000 galaxies in the range $0.005 < z < 0.2$ with the $\Lambda$LTB model, finding support for a deep void \citep[][]{Keenan:2013mfa}. However, the comparison with supernova data did not confirm this finding.
\citet{Cai:2020tpy} obtained similar results when comparing to supernova data.
Finally, \citet{Castello:2021uad} fitted the $\Lambda$LTB model to supernova and BAO data, together with a distance prior on the CMB. They also found that a local inhomogeneity cannot explain away the Hubble tension.
Our analysis improves on previous work by considering subsets of supernova data, a more comprehensive set of observations and by adopting an improved statistical analysis.

This paper is organized as follows. In Section~\ref{sec:LLTB} we briefly review the $\Lambda$LTB model and discuss how to estimate the Hubble constant in a inhomogeneous Universe, and in Section~\ref{sec:observables} we discuss the observations used to constrain the $\Lambda$LTB model. We then show our results in Section~\ref{sec:results} and discuss them in Section~\ref{sec:discussion}. We conclude in Section~\ref{sec:conclusions}.

\section{An inhomogeneous Universe} \label{sec:LLTB} 

In this section, we briefly review the $\Lambda$LTB model. We place the observer at the center of the inhomogeneous region, effectively neglecting anisotropic degrees of freedoms. We also discuss and propose three different ways to compute the Hubble constant in an inhomogeneous but isotropic Universe. Hereafter, we will use the prime to denote a partial derivative with respect to the radial coordinate, $r$, while the dot will be used to denote a partial derivative with respect to the time coordinate, $t$. Additionally, we set $c=1$.

\subsection{The $\Lambda$LTB model}

The Lema\^\i tre-Tolman-Bondi metric (LTB) can be written as \citep[see][for a comprehensive review]{Marra:2022ixf}:
\begin{equation}
ds^2 = -dt^2 + \frac{R'^2(r,t)}{ 1+2r^2k(r)\tilde{M}^2} dr^2 + R^2(r,t) d\Omega \,, \label{eq:metric}
\end{equation}
where $ d\Omega = d\theta^2 +\sin^2\theta d\phi^2$,  $\tilde{M}$ is an arbitrary mass scale and $k(r)$ is a free function. The Friedmann-Lema\^\i tre-Robertson-Walker metric (FLRW) can be recovered by imposing $k(r) = \mathrm{constant}$ and $R(r,t) = a(t) r$, with $a(t)$ being the FLRW scale factor.
Besides the curvature profile, $k(r)$, the $\Lambda$LTB model has two more arbitrary functions:  the mass function, $m(r)$, and the Big Bang time function, $t_{BB}(r)$. We set these functions following \citet{Camarena:2021mjr}, that is, we adopt a homogeneous Big Bang time $t_{BB}(r) = 0$ and set the radial coordinate gauge such that $m(r)=4\pi\tilde{M}^2 {r^3}/{3}$. A homogeneous Big Bang time is necessary to ensure the absence of large inhomogeneities at early times, in agreement with the standard paradigm of inflation \citep{Zibin:2008vj} 

After fixing $m(r)$ and $t_{\mathrm{BB}}(r)$, one is left with the curvature function $k(r)$, which we model according to
\begin{equation}
k(r) = k_b + (k_c - k_b) P_3(r/r_B,0) \,, \label{eq:kr}
\end{equation}
where $k_b$ and $k_c$ are the curvature outside and at the center of the spherical inhomogeneity, respectively, $r_B$ is the comoving radius of the inhomogeneity and the function $P_n(x)$ follows \citep{Valkenburg:2012td}:
\begin{align}
P_n (x) =  \left\lbrace 
\begin{array}{lll}
1-\exp \left [- \left (1- x \right)^n /x \right]  \phantom{ciao} & 0 \leq x < 1 \,, \label{eq:Pn} \\ 
 0   & 1 \leq x \,.
\end{array}
\right. 
\end{align}
This curvature profile describes a compensated spherical inhomogeneity, that is, our $\Lambda$LTB model simply becomes a $\Lambda$CDM model at  $r \geq r_B$. Furthermore, equation~\eqref{eq:kr} establishes the existence of $r_L$, the compensating scale, at which the central over/underdense region makes a transition to the surrounding mass-compensating under/overdense region.

Due to the radial dependence of $R(r,t)$, the expansion of the universe is not only inhomogeneous but also anisotropic. Then, there exists two scalar factors: the transverse one, $a_{\perp}(r,t) = R(r,t)/r $, and the longitudinal one, $a_{\parallel}(r,t) = R'(r,t)$. This means that there are also two expansion rates defined as
\begin{align}
H_{\perp}(r,t) & = \frac{\dot{a}_{\perp}(r,t)}{a_{\perp}(r,t)} \,, \label{eq:HubbleT} \\
H_{\parallel}(r,t)  & = \frac{\dot{a}_{\parallel}(r,t)}{a_{\parallel}(r,t)} \,. \label{eq:HubbleL}
\end{align}
Using the previous equations, we can define the present-day density parameters as:
\begin{align}
\Omega_{\Lambda ,0}(r) & = \frac{\Lambda}{3H_{\perp}^2(r,t_0)} \,,  \label{eq:OmegaL} \\
\Omega_{m,0}(r) & = \frac{2 m(r)}{R^3(r,t_0) H_{\perp}^2(r,t_0)} \,, \label{eq:Omegam} \\ 
\Omega_{k,0}(r) & = \frac{2 r^2 k(r) \tilde{M}^2}{R^2(r,t_0) H_{\perp}^2(r,t_0)} \,.  \label{eq:Omegak}
\end{align}
For the sake of simplicity, hereafter we use $a~\equiv~a_\perp$ and $H~\equiv~H_\perp$, unless otherwise stated.

The matter density contrast is defined by
\begin{equation}
\delta \rho (r,t) =  \frac{\rho_m (r,t)}{ \rho_m (r_B,t)} - 1 \label{eq:deltam} \,,
\end{equation}
and the mass (integrated) density contrast is given by
\begin{align}
\nonumber \delta (r, t_0) & = \frac{4\pi \int^r_0 d\overline{r} \delta \rho(\overline{r},t_0) R^2(\overline{r},t_0) R'(\overline{r},t_0)}{4\pi R^3(r,t_0)/3} \\
   & = \frac{\Omega_{m,0}(r) \, H_0^2(r)}{\Omega_{m,0}^{\rm{out}}\, {H_{0}^{\rm{out}}}^2} -1 \,,  \label{eq:Deltar}
\end{align}
where  $H_0(r) \equiv H(r,t_0)$. 
Hereupon, we use the superscript ``out'' to denote quantities outside the inhomogeneity, i.e. FLRW background quantities. We additionally define the FLRW comoving coordinate at the present time  as:
\begin{equation}
r^{\rm out} = R(r,t_0)/a^{\rm out}(t_0) \,. \label{eq:rout}
\end{equation}

The $\Lambda$LTB model is specified by the parameters that characterize the inhomogeneity, $r_B$ and $k_c$ in equation~\eqref{eq:kr}, and by the standard six $\Lambda$CDM parameters. The latter are the Hubble constant, the baryon density, the cold dark matter density, the optical depth, the amplitude of the power spectrum, and its tilt.
Regarding the $\Lambda$LTB parameters, instead of $r_B$ and $k_c$, we adopt $z_B$, which is the redshift corresponding to $r_B$, and $\delta_0$, which is the contrast at the center. The motivation for this change of independent variables is that  $z_B$ and $\delta_0$ are easier to interpret as far as the low-redshift universe is concerned, the subject of this paper.
In the following we will show our results using $r_L^{\mathrm{out}}$, the compensating scale in FLRW comoving coordinates, and $\delta_L \equiv \delta(r_L,t_0)$, the mass density contrast at the aforementioned scale \citep[for illustrative plots, see][]{Camarena:2021mjr}.
Finally, in order to improve the convergence of the Monte Carlo Markov Chain (MCMC), we normalize $\delta_0$ such that, instead of $-1 \leq \delta_0 < \infty$, we use:
\begin{align}
\tilde{\delta}_0  & = \left\lbrace 
\begin{array}{rl}
\delta_0 \phantom{ciaociao} & \delta_0 \leq 0\ \\
\delta_0/(1+\delta_0) \phantom{ciaociao} & \delta_0 > 0\ 
\end{array}
\right. \,,
\end{align}
which satisfies $-1\le \tilde \delta_0 <1$.
The same normalization is applied for the $\delta_L$. For the sake of simplicity, hereafter we drop the tilde.

\subsection{Anisotropies} \label{anisotropies}

As said earlier, we consider the observer at the center of a spherical structure,  a scenario in which observations are perturbed in a spherically symmetric way.
As the universe is both radially inhomogeneous and anisotropic, one may argue that an anisotropic perturbation of observations should be considered.
To this point one may consider a more general metric such as the quasi-spherical Szekeres model \citep{1975CMaPh..41...55S}, which features a dipole inhomogeneity instead of a spherical one \citep{Bolejko:2006my}, or simply displace the observer from the origin \citep[][]{Alnes:2006pf}.

Our modeling, however, is justified \textit{a priori} by the fact that we wish to understand if a local underdensity can explain away the Hubble tension.
Indeed, this calls for a 9\% increase in the local Hubble rate, which means that the observer must be within a deep underdensity of contrast $\approx -0.5$, see Eq.~\eqref{linH}, with subdominant anisotropic corrections.
The smaller axis of an underdense ellipsoid grows indeed faster as compared to the longer ones, with the consequence that voids become increasingly spherical as they evolve.
If then the observer is misplaced from the center of such a structure, they will develop a peculiar velocity with respect to the CMB of approximately $ v = \Delta H \, d_{\rm obs}$, where $\Delta H \simeq 6$ km/s/Mpc and $d_{\rm obs}$ is the distance from the center \citep[][]{Marra:2011ct}.
As the observed CMB dipole is $v/c \simeq 1.2 \times 10^{-3}$ \citep[][]{Planck:2018nkj}, this means that $d_{\rm obs} \lesssim 60$ Mpc, which is small as compared to the size of the inhomogeneity (see Fig.~\ref{fig:delta_alpha}): in the standard model a source at $z=0.15$, the maximum redshift considered in the local $H_0$ determination by SH0ES, is at a distance of $\approx 600$ Mpc.
We conclude that our modeling is adequate for testing the local-void scenario.
On the other hand, it is worth stressing that the local-void scenario fine-tunes the position of the observer by $\approx(60/600)^3=1/1000$ chances. In other words, if successful, one trades a one-in-a-million (5$\sigma$) inconsistency in data with a one-in-a-thousand fine-tuning.

\subsection{The Hubble constant} \label{sec:H0}

Although the background FLRW expansion is well defined by the value of $H_0^{\mathrm{out}}$, due to the radial dependency on the expansion rate, $H_0(r) \neq$ constant, our model does not possess a unique definition of the Hubble constant. In addition, there does not exist, \textit{a priori}, any preferable scale, $r_x$, at which one can safely define $H_0 = H_0(r_x)$ -- the definition of the Hubble constant remains arbitrary. 

Here, we use observational reasoning and extend FLRW concepts to propose three definitions of the Hubble constant for a $\Lambda$LTB universe. These approaches use a mock catalog of supernovae in the redshift range $0.023 < z < 0.15$, which is generated considering $\Lambda$LTB luminosity distances as the observed quantity, and the redshift distribution and covariance matrix of the Pantheon dataset \citep{Scolnic:2017caz}. This mock data set is generated at each sampled point of the parameter space in order to correctly account for the different cosmological model and it is used only for the determination of the predicted Hubble constant.

\begin{figure}
\centering
\includegraphics[width=\columnwidth]{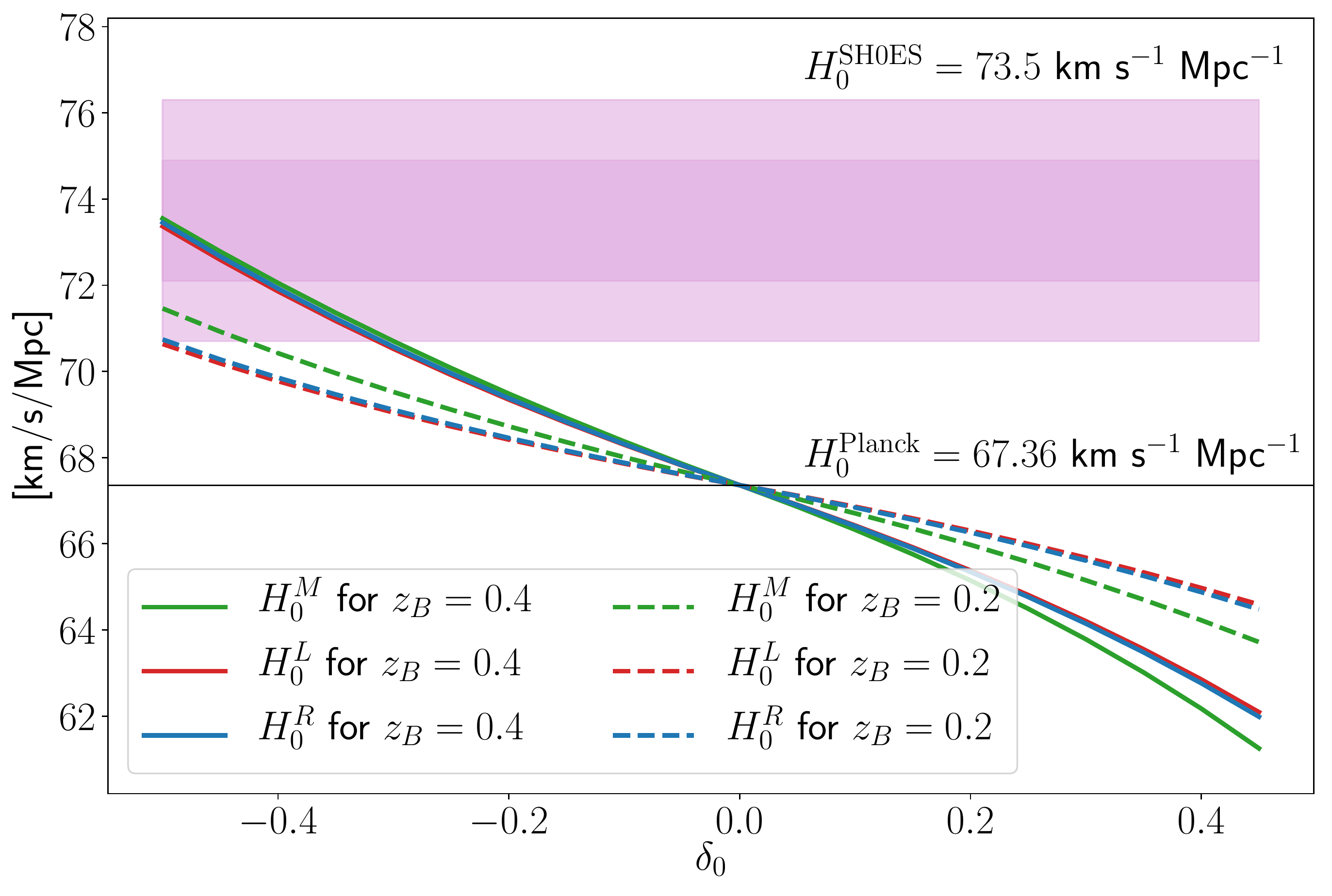}
\caption{$H_0^{\mathrm{M}}$, $H_0^{\mathrm{R}}$ and $H_0^{\mathrm{L}}$ as a function of the central contrast~$\delta_0$ for $z_B = 0.2$  and $z_B = 0.4$.
A local void  with $\delta_0 \approx -0.5$ and $z_B = 0.2$ or $\delta_0 \approx -0.3$ and $z_B = 0.4$ can potentially solve the Hubble crisis by providing a background expansion $H_0^{\mathrm{out}} = H_0^{\mathrm{Planck}}$ (horizontal black line) and a local rate that agrees with $ H_0^{\mathrm{SH0ES}}$ (pink region).  \label{fig:H0}}
\end{figure}

\subsubsection{Mean Hubble constant $H_0^{\mathrm{M}}$} \label{sec:HM}

Our first approach, dubbed as $H_0^{\mathrm{M}}$, is an extension of the one proposed in \citet{Valkenburg:2012td}. $H_0^{\mathrm{M}}$ is obtained from a weighted comparison between the luminosity distance and a radial dependent cosmographic expansion over the range $0.023 < z < 0.15$:
\begin{equation}\label{eq:H0V}
\frac{c}{H^M_0} = \int_{0.023}^{0.15} d_L(z) W(z)\left\lbrace z + \frac{1}{2} \left[1-q_0(r)\right]z^2 \right\rbrace^{-1} dz \,, 
\end{equation}
where $W(z)$ is the normalized redshift distribution of the mock supernovae and the deceleration parameter is defined as 
\begin{equation}\label{eq:q0r}
q_0(r) = \Omega_{m,0}(r)/2 - \Omega_{\Lambda ,0}(r) \,.
\end{equation}
%

\subsubsection{SH0ES Hubble constant $H_0^{\mathrm{R}}$} \label{sec:HR}

For our second definition we adopt the procedure proposed in \citet{Redlich:2014gga}, and lately in \citet{Efstathiou:2021ocp}, where the Hubble constant is obtained by mimicking the typical cosmic distance ladder procedure~\citep[see for instance][]{Riess:2016jrr}, i.e., fitting the mock catalog using the FLRW cosmographic expansion and assuming a constant deceleration parameter $q_0 = -0.55$ along with a constant value for $H_0$. 
This determination, dubbed $H_0^{\mathrm{R}}$, neglects any spatial degrees of freedom introduced by the LTB metric and it could be used to identify if deviations of  statistical homogeneity could substantially bias  cosmic distance ladder determinations.
We would like to stress that, while \citet{Redlich:2014gga} first presented this method in the context of inhomogeneous models, \citet{Efstathiou:2021ocp}  proposed this approach to point out that the cosmic distance ladder technique does not  determine the Hubble rate at $z= 0$ but in a specific low-$z$ range given by the set of supernovae that is adopted in the cosmic distance ladder.

\subsubsection{Local Hubble constant $H_0^{\mathrm{L}}$} \label{sec:HL}

Lastly, we propose $H_0^{\mathrm{L}}$, which is determined as $H_0^{\mathrm{R}}$ but with a radial dependent deceleration parameter:
\begin{equation}\label{eq:q0r2}
\tilde{q}_0(r,H_0^L) = q_0 (r) \left[\frac{H_0(r)}{H_0^L}\right]^2 \,,
\end{equation}
where the last factor enforces the constant $H_0^{\mathrm{L}}$ as the local Hubble rate when defining the density parameters of equations~(\ref{eq:OmegaL}-\ref{eq:Omegam}).

\vspace{.5 cm}

Figure~\ref{fig:H0} shows $H_0^{\mathrm{M}}$, $H_0^{\mathrm{R}}$ and $H_0^{\mathrm{L}}$ as a function of $\delta_0$ for two particular values of the boundary redshift: $z_B = 0.2$ (dashed lines) and $z_B = 0.4$ (solid lines). One can note that $H_0^{\mathrm{R}}$ (blue lines) and $H_0^{\mathrm{L}}$ (red lines) provide similar values for any pair of $\delta_0$ and $z_B$.
On the other hand, $H_0^{\mathrm{M}}$ (green lines) enhances the deviations from $H_0^{\mathrm{out}}$, especially at $|\delta_0| \gtrsim 0.1$. 
A local void  with $\delta_0 \approx -0.5$ and $z_B = 0.2$ or $\delta_0 \approx -0.3$ and $z_B = 0.4$ can potentially solve the Hubble crisis by providing a background expansion in agreement with the CMB, $H_0^{\mathrm{out}} = H_0^{\mathrm{Planck}}$, and a local rate that agrees with SH0ES.

\section{Observables} \label{sec:observables}

In order to constrain the $\Lambda$LTB model we use: Planck 2018 data coming from the high-$\ell$ and low-$\ell$ TT+TE+EE power spectrum \citep{Aghanim:2018eyx}; BAO measurements from 6dFGS \citep{Beutler:2011hx}, SDSS-MGS \citep{Ross:2014qpa} and BOSS-DR12 \citep{Alam:2016hwk}; cosmic chronometers data from \cite{Moresco:2016mzx,Moresco:2012jh,Simon:2004tf,Stern:2009ep,Zhang:2012mp,Moresco:2015cya};%
\footnote{See \citet{Moresco:2022phi} for the most recent compilation.}
type Ia supernovae distances from Pantheon compilation \citep{Scolnic:2017caz}; a $2\sigma$ upper limit prior on the Compton $y$-distortion provided by COBE-FIRAS \citep{Fixsen:1996nj}; a prior on the amplitude of kSZ effect at $\ell=3000$ \citep{Reichardt:2020jrr}; and the  Cepheid calibration of the absolute magnitude of supernovae, $M_B$, from \citet{Camarena:2019moy,Camarena:2021jlr}. See \cite{Camarena:2021mjr} for a thorough discussion of this data and its $\Lambda$LTB theoretical description. 

We will carry out our analyses using several combinations of the aforementioned data.
Moreover, we will also consider combinations of data including not the whole set of Pantheon supernovae but only supernovae  in the redshift range $0.023 < z <0.15$---the so-called Hubble flow supernovae that are used by SH0ES in the determination of $H_0$. We dub this subset of the Pantheon catalog as low-$z$ supernovae. Additionally, we carried out analyses including a prior on the Hubble constant, instead of a prior on $M_B$. Specifically, we impose the SH0ES determination $H_0 = 73.5 \pm 1.4$~km~s$^{-1}$~Mpc$^{-1}$ \citep{Reid:2019tiq} on $H_0^{\mathrm{L}}$.
The aim of these extra analyses is to demonstrate that both methods, either with a prior on $M_B$ or a prior on $H_0$, are statistically equivalent when the local $H_0$ prior is implemented considering that the cosmic distance ladder technique does not measure the Hubble rate at $z=0$ but rather in a specific redshift range \citep{Efstathiou:2021ocp}.

\section{Results} \label{sec:results}

\begin{figure}
\centering
\includegraphics[width=\columnwidth]{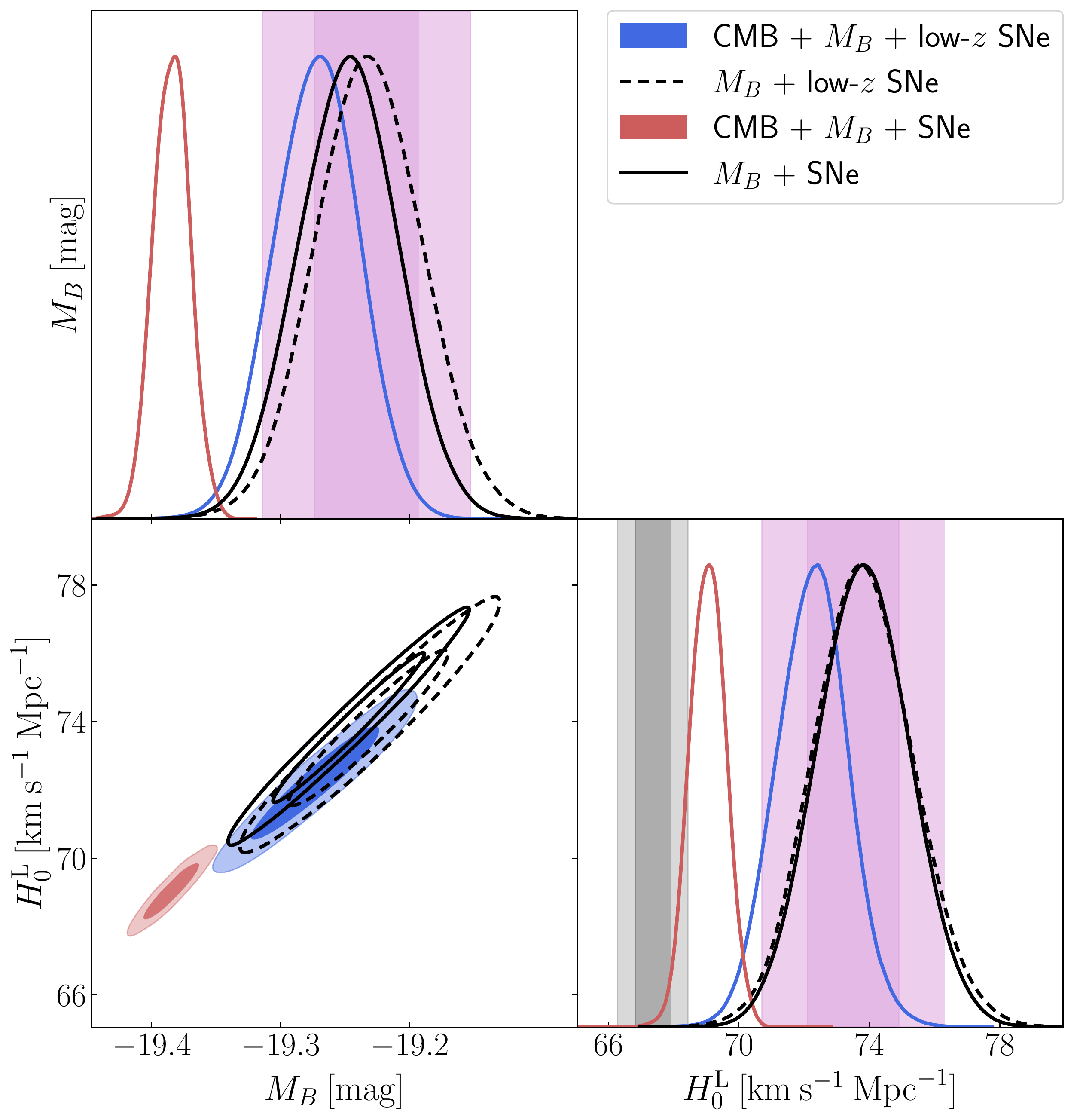}
\includegraphics[width=.95 \columnwidth]{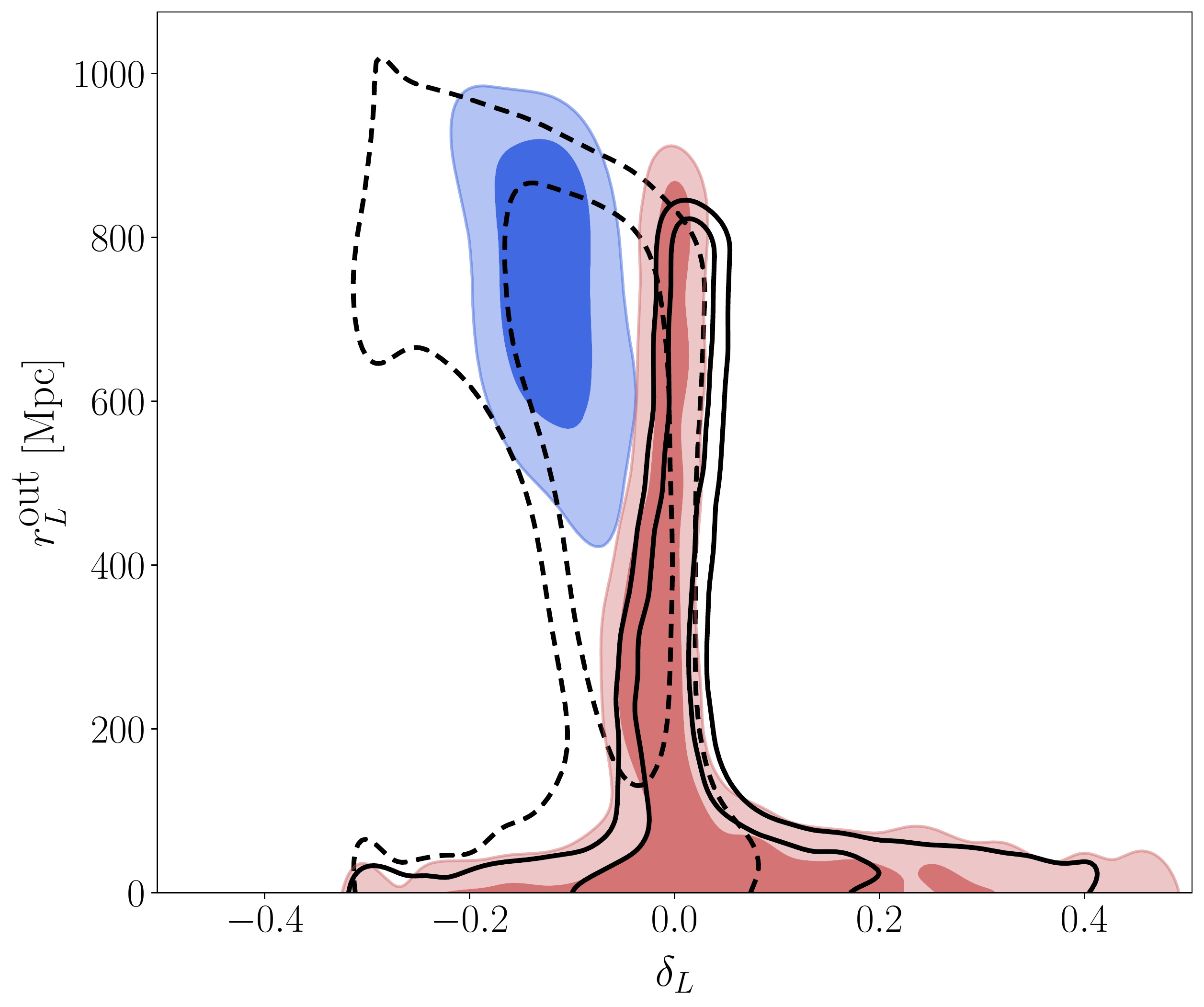}
\includegraphics[trim = 0. 0. 0. 3cm, clip, width= \columnwidth]{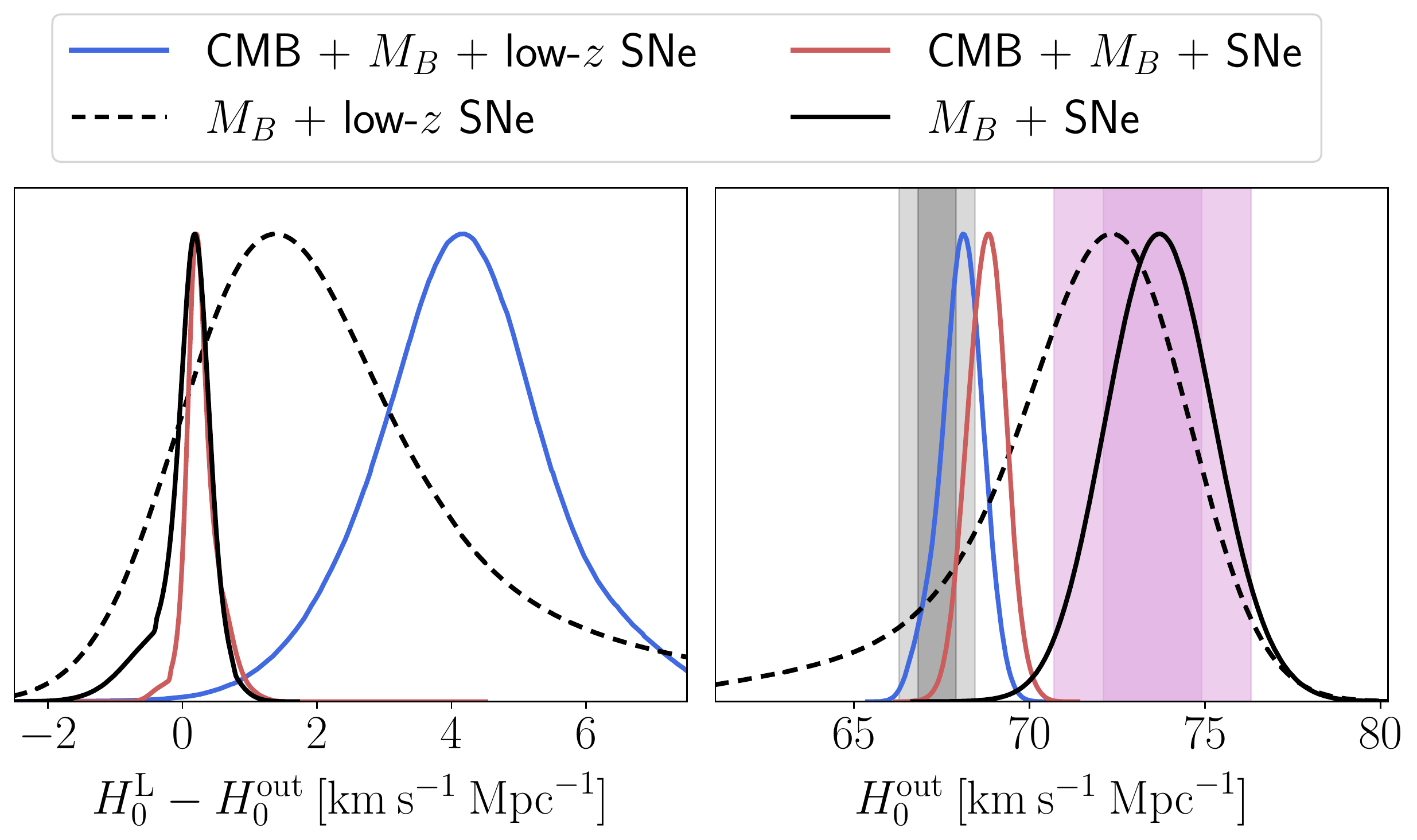}
\caption{Marginalized constraints, at 68\% and 95\% confidence level, on several parameters of interest when considering, in a flat background Universe, combinations of CMB and supernova data, together with the local prior on the supernova absolute magnitude $M_B$.
Shown are $M_B$ and the local Hubble rate $H_0^{\rm L}$ (top), the effective mass density contrast $\delta_L$ and compensating scale $r_L^{\rm out}$ of the $\Lambda$LTB model (center), and background Hubble constant $H_0^{\rm out}$ and the local increase with respect to the background rate, $\Delta H =H_0^{\rm L}- H_0^{\rm out} $ (bottom). Note that there is tension only when considering all supernovae and the CMB. See Section~\ref{sub:flat}.
\label{fig:LLTB_all_rL}}
\end{figure}

\begin{figure*}
\centering
\includegraphics[width=\textwidth]{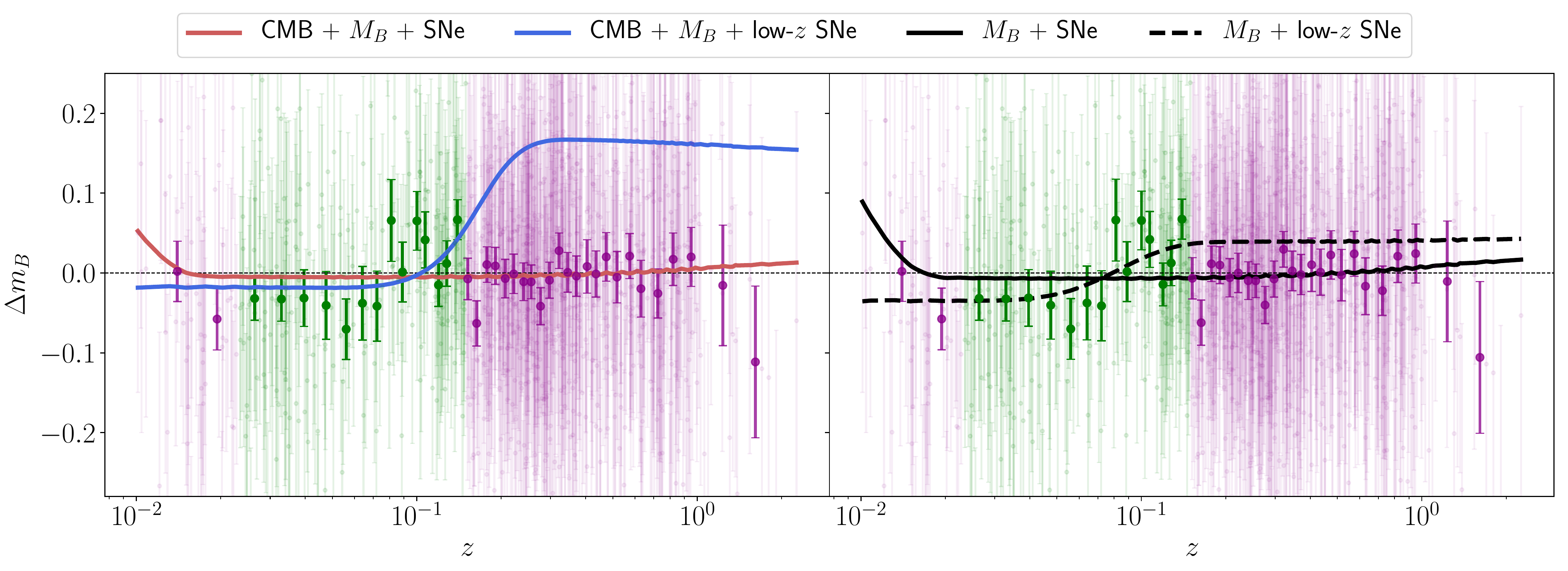}
\caption{Apparent magnitude residuals of the Pantheon supernovae, as function of the redshift, taking as a reference the best fit of the $\Lambda$CDM model to the combination CMB~+ $M_B$~+~SNe + All. One can see, from the left panel, that the best fit of the $\Lambda$LTB model to CMB + $M_B$ + low-$z$ SNe (blue line) fits  well the supernovae in the range $0.023<z<0.15$ (green data points) and provides a solution to the Hubble crisis, see Section~\ref{sub:flat}.
However,  the other supernovae (purple data points) constrain the $\Lambda$LTB luminosity distance (red line) to a shape similar to the $\Lambda$CDM one. The result is that the $\Lambda$LTB model cannot explain the Hubble tension.
The right panel shows the case without CMB data. While the full supernova sample does not prefer an underdensity (solid curve), when only considering low-$z$ supernovae one sees that the profile is compatible with a local void (dashed black line). This is due to a fluctuation in the supernova apparent magnitudes at $0.1\lesssim z \lesssim 0.15$.
\label{fig:BF_SNe}}
\end{figure*}

Data analysis is performed using the \textsc{montelltb} code \citep{Camarena:2021mjr}, which conveniently wraps a modified version of the $\Lambda$LTB solver \textsc{vd2020} \citep{Valkenburg:2011tm} in \textsc{montepython} \citep{Brinckmann:2018cvx,Audren:2012wb}. We explore via MCMC the parameter space, and evaluate the convergence of our chains demanding $(R-1) \lesssim 0.05 $ for the inhomogeneous parameters and  $(R-1) \sim \mathcal{O}(10^{-3})$ for the $\Lambda$CDM background parameters, where $R$ is the Gelman-Rubin diagnostic \citep{Gelman:1992zz}. Most of the plots displayed in this section were generated using \textsc{getdist} \citep{Lewis:2019xzd}.

We extend the assumptions made in \cite{Camarena:2021mjr} and  consider both a flat and a curved $\Lambda$CDM background. Given that Planck data has showed a moderate evidence for a closed Universe \citep{DiValentino:2019qzk,Handley:2019tkm}, the question if our Universe is flat or curved has been  recently investigated \citep[see,e.g.,][]{Vagnozzi:2020dfn,Vagnozzi:2020rcz}. Additionally, the FLRW curvature has been found to have a strong correlation with a possible change in the CMB temperature, potentially pointing out the existence of a strong correlation with the parameters of an inhomogeneous model \citep{Bose:2020cjb,Ivanov:2020mfr}.

As mentioned before, we considered several combinations of the data discussed in Section~\ref{sec:observables}. We denote as Base the combination of CMB, SNe and the local prior (either on $H_0$ or $M_B$). We quantify the tension on $H_0$ and $M_B$ assuming the one-dimensional Gaussian limit of the index of inconsistency; a moment-based estimator that can be use to quantitatively measure discordance~\citep{Lin:2017ikq}.

\subsection{Flat background FLRW metric} \label{sub:flat}

We start by considering a flat $\Lambda$CDM background ($\Omega_k^{\mathrm{out}} = 0$) and only CMB and supernova observations, together with the local prior on the supernova absolute magnitude $M_B$.
Figure~\ref{fig:LLTB_all_rL} shows marginalized constraints on several parameters of interest for four observable combinations.\footnote{See Appendix~\ref{ap:H-M} for the plot relative to the case with the local prior on $H_0$.}
Figure~\ref{fig:BF_SNe} shows the corresponding apparent magnitude residuals  of the $\Lambda$LTB best fits with respect to the $\Lambda$CDM best fit.

As it is well known, the freedom in defining the LTB curvature function allows one to fit any luminosity-distance-redshift relation, that is, any supernova sample. If one adds a prior on $M_B$, then the latter simply constrains the supernova absolute magnitude, and so local $H_0$, without changing the fit to supernova data. We start by discussing this case for the full Pantheon sample and its low-redshift subset ($0.023 < z <0.15$). From Figure~\ref{fig:LLTB_all_rL} we see that the constraints on $\delta_L$ and $r_L^{\rm out}$ from the full SN sample (solid black lines) are along the $\delta_L=0$ axis, not favoring under- or overdensities. In particular, one has $H_0^{\rm L}\approx H_0^{\rm out} \approx H_0^{\mathrm{SH0ES}}$.
In other words, there is no local void nor $H_0$ tension, as expected.

If one considers only low-$z$ supernovae (dashed black lines), the situation is qualitatively the same, albeit with weaker constraints. Note, however, that a local underdensity is somewhat preferred: this is caused by a fluctuations in the supernova apparent magnitudes at $0.1\lesssim z \lesssim 0.15$, as evident from Figure~\ref{fig:BF_SNe}. Because of this allegedly random fluctuation, there is a small shift between $H_0^{\rm L}$ and $H_0^{\rm out}$, see Figure~\ref{fig:LLTB_all_rL}.

Next we add CMB observations, which are fit by a lower background $H_0$ as compared with $H_0^{\mathrm{SH0ES}}$.
If we consider low-$z$ supernovae (blue curves), then one can have all the supernovae inside a local underdensity and is free to fit any $\Delta H =H_0^{\rm L}- H_0^{\rm out} $, see Figures~\ref{fig:LLTB_all_rL} and~\ref{fig:BF_SNe}.
Specifically, the data favors a local underdensity and the local value of the Hubble rate is in agreement with the local prior and the tension between CMB observations and the local prior is solved.
Note also that the local calibration of $M_B$ is not affected by CMB observations.
Table~\ref{tab:LLTB_params} shows the marginalized constraints for the relevant  parameters, including $H_0^{\mathrm{L}}$, $H_0^{\mathrm{R}}$, and $H_0^{\mathrm{M}}$.
We also show the change in the observed CMB temperature $\Delta T \equiv T_0^{\mathrm{obs}} - T_0^{\mathrm{out}}$, with $T_0^{\mathrm{obs}}$ being the CMB temperature measured by the observer  and $ T_0^{\mathrm{out}} = 2.7255$~K the background temperature.\footnote{Note that we have neglected possible dynamical effects of radiation \citep{Clarkson:2010ej}}
Indeed, analogous to other parameters, the observer at the center of the inhomogeneity is expected to measure a different CMB temperature as compared with the expected FLRW background temperature. This change in the temperature is strongly related to the features of the inhomogeneity.
Within this scenario one expects a $\approx 2$ mK change in the CMB temperature.
It is worth pointing out that the fact that the analysis $M_B$ + low-$z$ SNe also suggests a similar underdensity is a coincidence: even without the fluctuation at $0.1\lesssim z \lesssim 0.15$ one would have obtained here a similar result.

Then, we consider the full Pantheon sample. In this case, the luminosity-distance-redshift relation mapped by the supernovae does not allow for a sufficiently large and deep underdensity that can solve the $H_0$ tension: a sudden change in the luminosity distance is not allowed by the supernovae at $z>0.15$, see Figures~\ref{fig:LLTB_all_rL} and~\ref{fig:BF_SNe}. In particular, CMB data induce a lower value of $M_B$, at odds with the local prior, the so-called $M_B$ tension \citep[][]{Camarena:2021jlr}.
Also, in this case, the change in the CMB temperature is much smaller, approximately $\approx 0.01$ mK.
Our results are that a local void is not favored by the data and the $H_0$ tension is not solved.
Note, however, that $\Delta H =H_0^{\rm L}- H_0^{\rm out} $ does prefer small but positive values, that is, and underdensity. We will come back to this in Section~\ref{sub:local}.

Finally, we include other observables, considering all the combinations discussed in \citet{Camarena:2021mjr}. Table~\ref{tab:LLTB_params} presents the relevant results, including the corresponding $\chi^2_{\mathrm{min}}$ and the resulting tensions on $M_B$ and $H_0$, with respect to \citet{Camarena:2019moy} and \citet{Reid:2019tiq}, respectively.

\begin{table*}
\begin{center}
\setlength{\tabcolsep}{5pt}
\renewcommand{\arraystretch}{1.55}
\caption{$68\%$ confidence level intervals for the relevant parameters for the different combinations of data here analyzed, considering both the prior on $H_0$ and $M_B$. We also report the  $\chi^2_{\mathrm{min}}$ and the tensions on $H_0$ and $M_B$ in sigma units.\label{tab:LLTB_params}}
\begin{tabular}{ccccccc}
\hline 
\hline
Parameter & CMB + loc. prior + low-$z$ SNe & Base & Base + BAO + Hz & Base + $y$-dist. & Base + kSZ & Base + All \\ \hline
\hline
\multicolumn{7}{c}{Prior on $M_B$} \\ \hline
$M_B$ [mag] & $-19.271^{+0.032}_{-0.035}$ & $-19.384^{+0.014}_{-0.014}$ & $-19.389^{+0.011}_{-0.012}$ & $-19.386^{+0.014}_{-0.014}$ & $-19.384^{+0.014}_{-0.014}$ & $-19.391^{+0.012}_{-0.012}$ \\ \hline
$H_0^{\mathrm{M}}$ [km/s/Mpc] & $72.47^{+1.09}_{-1.10}$ & $69.06^{+0.53}_{-0.60}$ & $68.89^{+0.44}_{-0.46}$ & $69.01^{+0.58}_{-0.55}$ & $69.07^{+0.56}_{-0.54}$ & $68.77^{+0.40}_{-0.46}$ \\ \hline
$H_0^{\mathrm{L}}$ [km/s/Mpc] & $72.29^{+1.11}_{-1.12}$ & $69.06^{+0.54}_{-0.57}$ & $68.89^{+0.43}_{-0.46}$ & $69.00^{+0.57}_{-0.53}$ & $69.07^{+0.55}_{-0.54}$ & $68.78^{+0.39}_{-0.44}$ \\ \hline
$H_0^{\mathrm{R}}$ [km/s/Mpc] & $72.38^{+1.12}_{-1.14}$ & $69.06^{+0.52}_{-0.54}$ & $68.90^{+0.41}_{-0.44}$ & $69.00^{+0.54}_{-0.51}$ & $69.07^{+0.53}_{-0.50}$ & $68.79^{+0.37}_{-0.42}$ \\ \hline
$\Delta T\ [\rm{m K}]$ & $1.861^{+0.639}_{-0.918}$ & $-0.017^{+0.042}_{-0.041}$ & $-0.007^{+0.027}_{-0.045}$ & $-0.009^{+0.025}_{-0.048}$ & $-0.023^{+0.023}_{-0.027}$ & $-0.022^{+0.022}_{-0.025}$ \\ \hline
Tension on $H_0$ & $0.7$ & $3.0$ & $3.1$ & $3.0$ & $2.9$ & $3.2$ \\ \hline
Tension on $M_B$ & $0.7$ & $3.5$ & $3.7$ & $3.6$ & $3.5$ & $3.8$ \\ \hline
$\chi^2_{\rm{min}}$ & $2996.1$ & $3808.7$ & $3826.7$ & $3807.3$ & $3808.0$ & $3828.2$ \\ \hline
\hline
\multicolumn{7}{c}{Prior on $H_0$} \\ \hline
$M_B$ [mag] & $-19.258^{+0.047}_{-0.044}$ & $-19.386^{+0.015}_{-0.015}$ & $-19.391^{+0.012}_{-0.012}$ & $-19.389^{+0.015}_{-0.015}$ & $-19.389^{+0.014}_{-0.014}$ & $-19.392^{+0.011}_{-0.012}$ \\ \hline
$H_0^{\mathrm{M}}$ [km/s/Mpc] & $73.01^{+1.49}_{-1.53}$ & $69.09^{+0.56}_{-0.58}$ & $68.90^{+0.47}_{-0.49}$ & $68.98^{+0.59}_{-0.60}$ & $68.94^{+0.55}_{-0.51}$ & $68.83^{+0.43}_{-0.46}$ \\ \hline
$H_0^{\mathrm{L}}$ [km/s/Mpc] & $72.83^{+1.53}_{-1.49}$ & $69.08^{+0.57}_{-0.56}$ & $68.88^{+0.44}_{-0.47}$ & $68.97^{+0.55}_{-0.58}$ & $68.94^{+0.54}_{-0.51}$ & $68.83^{+0.41}_{-0.44}$ \\ \hline
$H_0^{\mathrm{R}}$ [km/s/Mpc] & $72.94^{+1.56}_{-1.51}$ & $69.08^{+0.54}_{-0.53}$ & $68.89^{+0.46}_{-0.46}$ & $68.98^{+0.54}_{-0.55}$ & $68.94^{+0.51}_{-0.48}$ & $68.84^{+0.39}_{-0.42}$ \\ \hline
$\Delta T\ [\rm{m K}]$ & $2.145^{+0.869}_{-1.043}$ & $0.004^{+0.021}_{-0.065}$ & $0.007^{+0.024}_{-0.061}$ & $0.001^{+0.056}_{-0.067}$ & $-0.017^{+0.026}_{-0.032}$ & $-0.015^{+0.020}_{-0.036}$ \\ \hline
Tension on $H_0$ & $0.3$ & $2.9$ & $3.1$ & $3.0$ & $3.0$ & $3.2$ \\ \hline
Tension on $M_B$ & $0.4$ & $3.6$ & $3.8$ & $3.6$ & $3.6$ & $3.8$ \\ \hline
$\chi^2_{\rm{min}}$ & $2998.3$ & $3803.8$ & $3826.8$ & $3805.0$ & $3803.0$ & $3824.5$ \\ \hline
\end{tabular}
\end{center}
\end{table*}

\subsection{Curved background FLRW metric} \label{sub:curvature}

\begin{figure*}
\centering
\includegraphics[width=0.8\textwidth]{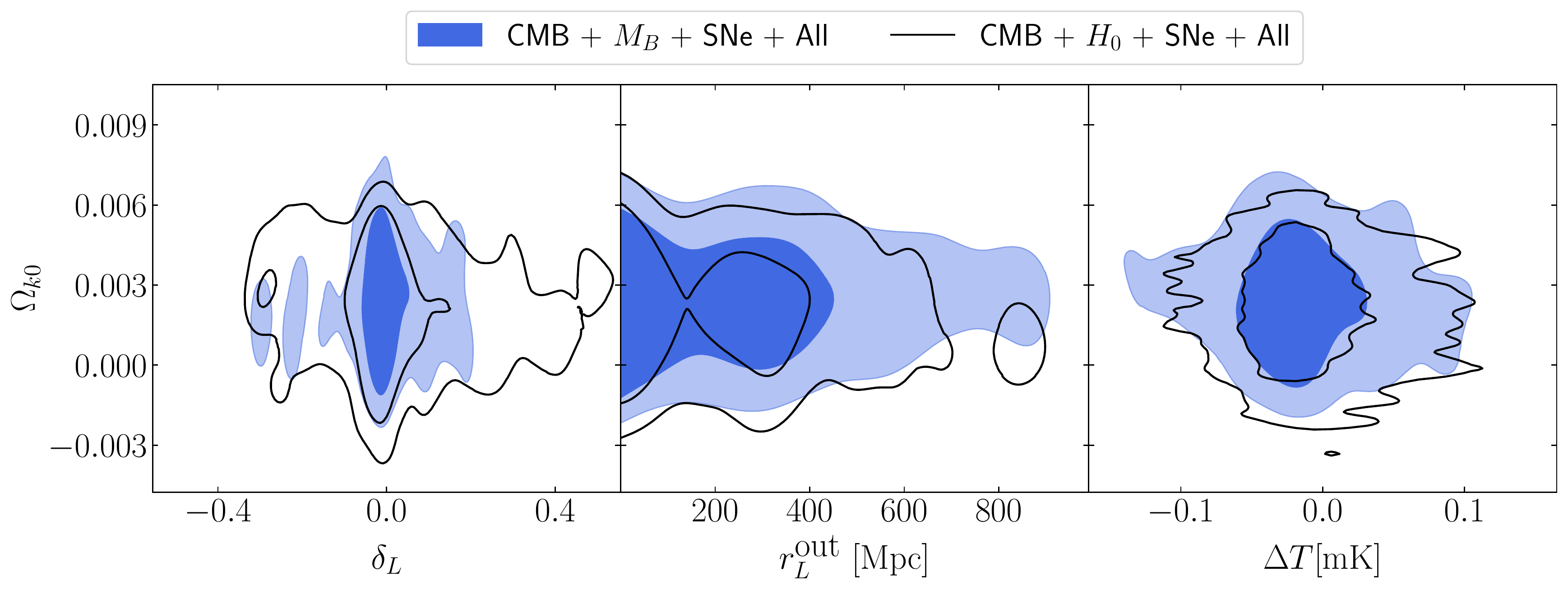}
\caption{Marginalized constraints on the effective mass density contrast $\delta_L$, compensating scale $r_L^{\rm out}$, temperature deviation $\Delta T$ and background curvature $\Omega_{k,0}$ at 68\% and 95\% confidence levels.
\label{fig:okLLTB_all_rL}}
\end{figure*}

\begin{table}
\begin{center}
\setlength{\tabcolsep}{5pt}
\renewcommand{\arraystretch}{1.55}
\caption{As Table~\ref{tab:LLTB_params}, but for a curved  background, $\Omega_{k,0} \neq 0$.\label{tab:okLLTB_params}}
\begin{tabular}{ccc}
\hline
\hline
Parameter & Base + BAO + Hz & Base + All \\ \hline
\hline
\multicolumn{3}{c}{Prior on $M_B$} \\ \hline
$\Omega_{k 0}$ & $0.0024^{+0.0016}_{-0.0016}$ & $0.0024^{+0.0017}_{-0.0017}$ \\ \hline
$M_B$ [mag] & $-19.372^{+0.016}_{-0.016}$ & $-19.373^{+0.017}_{-0.015}$ \\ \hline
$H_0^{\mathrm{M}}$ [km/s/Mpc] & $69.41^{+0.59}_{-0.61}$ & $69.38^{+0.62}_{-0.55}$ \\ \hline
$H_0^{\mathrm{L}}$ [km/s/Mpc] & $69.42^{+0.58}_{-0.59}$ & $69.38^{+0.61}_{-0.53}$ \\ \hline
$H_0^{\mathrm{R}}$ [km/s/Mpc] & $69.42^{+0.56}_{-0.58}$ & $69.39^{+0.59}_{-0.52}$ \\ \hline
$\Delta T\ [\rm{m K}]$ & $-0.016^{+0.026}_{-0.039}$ & $-0.020^{+0.020}_{-0.032}$ \\ \hline
Tension on $H_0$ & $2.7$ & $2.7$ \\ \hline
Tension on $M_B$ & $3.2$ & $3.2$ \\ \hline
$\chi^2_{\rm{min}}$ & $3828.4$ & $3829.0$ \\ \hline
\hline
\multicolumn{3}{c}{Prior on $H_0$} \\ \hline
$\Omega_{k 0}$ & $0.0022^{+0.0017}_{-0.0017}$ & $0.0021^{+0.0018}_{-0.0017}$ \\ \hline
$M_B$ [mag] & $-19.375^{+0.017}_{-0.017}$ & $-19.377^{+0.018}_{-0.015}$ \\ \hline
$H_0^{\mathrm{M}}$ [km/s/Mpc] & $69.43^{+0.61}_{-0.65}$ & $69.30^{+0.60}_{-0.57}$ \\ \hline
$H_0^{\mathrm{L}}$ [km/s/Mpc] & $69.41^{+0.59}_{-0.62}$ & $69.30^{+0.57}_{-0.56}$ \\ \hline
$H_0^{\mathrm{L}}$ [km/s/Mpc] & $69.43^{+0.58}_{-0.61}$ & $69.31^{+0.57}_{-0.55}$ \\ \hline
$\Delta T\ [\rm{m K}]$ & $0.004^{+0.028}_{-0.060}$ & $-0.014^{+0.016}_{-0.032}$ \\ \hline
Tension on $H_0$ & $2.7$ & $2.8$ \\ \hline
Tension on $M_B$ & $3.2$ & $3.3$ \\ \hline
$\chi^2_{\rm{min}}$ & $3820.1$ & $3822.7$ \\ \hline
\end{tabular}
\end{center}
\end{table}

We also study the case of a non-flat FLRW background. Results for these analyses are shown in Table~\ref{tab:okLLTB_params} and Figure~\ref{fig:okLLTB_all_rL}.
From Table~\ref{tab:okLLTB_params}, we can see that the inclusion of the curvature does not significantly change the overall results. In particular, the data favors a slightly open universe with  $\Omega_{k,0} \approx  0.002$, compatible with the flat case at $2\sigma$.
In particular, in Figure~\ref{fig:okLLTB_all_rL} we do not observe a strong correlation between $\Omega_{k,0}$ and the other parameters, in particular $\Delta T$, which remains constrained around zero.

\vspace{0.5cm}

\begin{figure}
\centering
\includegraphics[width=\columnwidth]{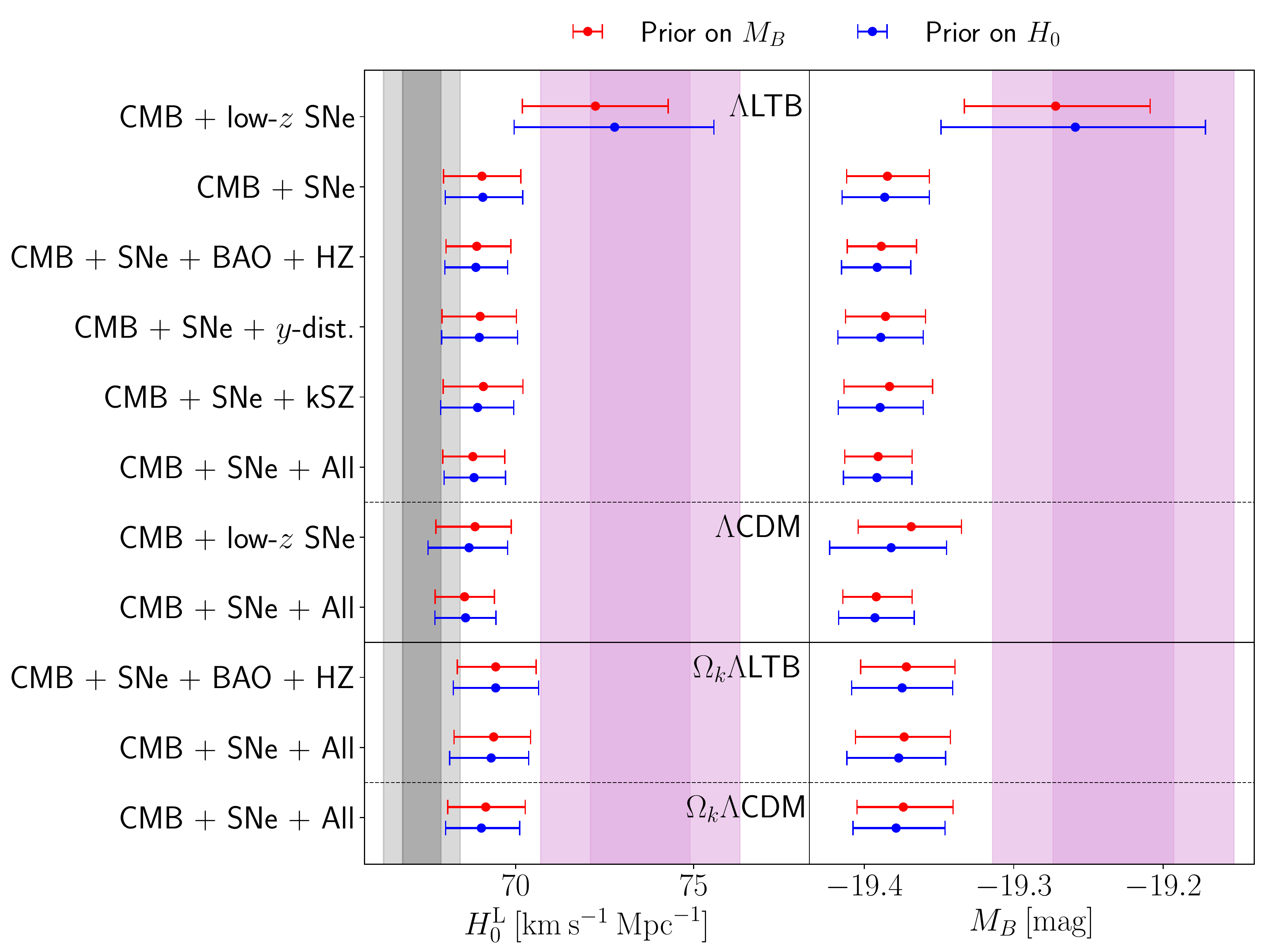}
\caption{Constraints on $H_0^{\mathrm{L}}$ and $M_B$ at 95\% confidence level for the cases here considered. The gray area corresponds to the value of the Hubble constant at 68\% and 95\% confidence level inferred from the CMB observations \citep{Aghanim:2018eyx}, while the pink areas correspond to the $H_0$ determination by SH0ES \citep{Reid:2019tiq} and the corresponding calibration of $M_B$ \citep[][]{Camarena:2019moy}.\label{fig:whisker}}
\end{figure}

Finally, Figure~\ref{fig:whisker} shows the different values obtained for $H_0^{\mathrm{L}}$ and $M_B$ for our different analyses, both considering a prior on $M_B$ and $H_0$. For the sake of the comparison, we have also included the results coming from analyses of the $\Lambda$CDM model.
We can see how the $\Lambda$LTB results follow the ones relative to the $\Lambda$CDM model.

\section{Discussion} \label{sec:discussion}

\subsection{Model selection}

\begin{table*}
\begin{center}
\setlength{\tabcolsep}{7pt}
\renewcommand{\arraystretch}{1.15}
\caption{
Results of the model selection analysis for the case of a flat background Universe ($\Delta x = x_{\Lambda\mathrm{LTB}} - x_{\Lambda\mathrm{CDM}}$). \label{tab:B01_flat}}
\begin{tabular}{ccccc}
\hline
\hline
Criteria & CMB + $M_B$ + low-$z$ SNe & CMB + $H_0$ + low-$z$ SNe & CMB + $M_B$ + All & CMB + $H_0$ + All \\ \hline
$\chi^2_{\Lambda\rm{CDM}}$ & 3014.3 & 3015.0 &  3830.0 & 3825.2 \\
$\Delta \chi^2$ & -18.2 & -16.7 & -1.8 & -0.7 \\
$\Delta$AIC  & -14.2 & -12.7 &  2.2 & 3.3 \\ 
$\ln B_{01}$ & -12.5 & -17.3 & 3.0 & 2.2 \\  
\hline
\end{tabular}
\end{center}
\end{table*}

\begin{table}
\begin{center}
\setlength{\tabcolsep}{7pt}
\renewcommand{\arraystretch}{1.15}
\caption{
Results of the model selection analysis for the case of a curved background Universe ($\Delta x = x_{\Lambda\mathrm{LTB}} - x_{\Lambda\mathrm{CDM}}$). \label{tab:B01_curved}}
\begin{tabular}{ccc}
\hline
\hline
Criteria & CMB + $M_B$ + All & CMB + $H_0$ + All \\ \hline
$\chi^2_{\Lambda\rm{CDM}}$ & 3828.7 & 3824.9 \\
$\Delta \chi^2$ & 0.4 & -2.2 \\
$\Delta$AIC & 4.4 & 1.8 \\ 
$\ln B_{01}$ & 2.2 & 2.4 \\  
\hline
\end{tabular}
\end{center}
\end{table}

\begin{table}
\begin{center}
\setlength{\tabcolsep}{5pt}
\renewcommand{\arraystretch}{1.55}
\caption{$68\%$ confidence level intervals for the relevant parameters. See Section~\ref{sub:curvature2} for details.\label{tab:alpha}}
\begin{tabular}{ccc}
\hline
\hline
Parameter & $\alpha$ free & $\alpha = 0$ \\ \hline
$M_B$ & $-19.372^{+0.016}_{-0.016}$ & $-19.391^{+0.012}_{-0.012}$ \\ \hline
$H_0^{\mathrm{M}}$ & $69.41^{+0.59}_{-0.61}$ & $68.77^{+0.40}_{-0.46}$ \\ \hline
$H_0^{\mathrm{L}}$ & $69.42^{+0.58}_{-0.59}$ & $68.78^{+0.39}_{-0.44}$ \\ \hline
$H_0^{\mathrm{R}}$ & $69.42^{+0.56}_{-0.58}$ & $68.79^{+0.37}_{-0.42}$ \\ \hline
$\Delta T [\rm{m K}]$ & $-0.016^{+0.026}_{-0.039}$ & $-0.022^{+0.022}_{-0.025}$ \\ \hline
Tension $H_0$ & $2.7$ & $3.2$ \\ \hline
Tension $M_B$ & $3.2$ & $3.8$ \\ \hline
$\chi^2_{\rm{min}}$ & $ 3827.9 $ & $3828.2$ \\ \hline
\end{tabular}
\end{center}
\end{table}

We have seen how the Hubble tension is solved when only low-redshift supernovae are considered but it is no longer solved when all supernovae are included. Here, we will quantify this statement using Bayesian model comparison between the $\Lambda$CDM and $\Lambda$LTB models.
We perform model selection  using the Bayes ratio. Since the $\Lambda$CDM model is nested in the $\Lambda$LTB model, we can  simplify the computation of the Bayes ratio by using the Savage-Dickey density ratio (SDRR) \citep{Trotta:2008qt}. This technique reduces the Bayes ratio to:
\begin{equation} \label{eq:B01}
B_{01} = \left. \frac{\int \mathcal{P}(\delta_0,z_B,\theta_i) \mathrm{d}\theta_i}{p(\delta_0)p(z_B)} \right|_{\delta_0=0,z_B=0} \,,
\end{equation}
with $\mathcal{P}$ being the posterior of the $\Lambda$LTB model, $\theta_i$ the $\Lambda$CDM background parameters, and $p$ the prior function. Although the SDRR can be safely applied to nested models, one should bear in mind that equation~\eqref{eq:B01} assumes that priors are separable, i.e., $p(\delta_0,z_B,\theta_i) = p(\delta_0)p(z_B)p(\theta_i)$.
Here, this assumption is fully satisfied since our analyses use wide flat priors over all parameters.\footnote{Except for $H_0^{\mathrm{L}}$ and $M_B$, but the priors are still separable.}
Specifically, we impose $z_B \in [0,0.5]$ and $\delta_0 \in [-1,1]$ such that the flat priors result in $p(\delta_0) = 1/2$ and $p(z_B) = 2$. 
In equation~\eqref{eq:B01} it is $B_{01} \propto \mathcal{E}_0/\mathcal{E}_1$, with $0$ representing the nested model, in our case the $\Lambda$CDM model, and $1$ the more complex model, the $\Lambda$LTB model. We qualitatively interpret the ratio $B_{01}$ via the Jeffreys' scale \citep{jeffreys1961theory}. Specifically, we adopt the conservative version discussed in \citet{Trotta:2008qt}, see Table~\ref{tab:jeffreys}.

We also use the Akaike information criterion (AIC):
\begin{equation}
\mathrm{AIC} = \chi^2_{\mathrm{min}} + 2 k \label{eq:aic}\,,
\end{equation}
with $k$ being the number of free parameters. The relative differences $\Delta \mathrm{AIC} \equiv \mathrm{AIC}_{\Lambda\mathrm{LTB}} - \mathrm{AIC}_{\Lambda\mathrm{CDM}}$ are qualitatively interpreted using the calibrated Jeffreys' scale shown in Table~\ref{tab:AIC}.

Results are shown in Tables~\ref{tab:B01_flat}~and~\ref{tab:B01_curved} for the flat and curved $\Lambda$CDM background, respectively. 
Under  the assumption of a flat background metric, we find a strong evidence, $B_{01} = -12.5$, in favor of the $\Lambda$LTB model when the CMB + $M_B$ + low-$z$ SNe data is considered.
This is confirmed by the $\Delta$AIC which shows no support to the $\Lambda$CDM model.
On the other hand, the inclusion of the full supernova dataset removes the preference for the $\Lambda$LTB model.
The analysis relative to the combination CMB + $M_B$ + All shows a moderate evidence for the $\Lambda$CDM model, $B_{01} = 3.0$, and a substantial support to the same model, $\Delta \textrm{AIC} = 2.2$. Similar results are obtained by considering a prior on $H_0$.
Finally, as can be seen from Table~\ref{tab:B01_curved}, the introduction of a non-vanishing background curvature does not qualitatively change the results  discussed above.

\subsection{Generalized curvature profile} \label{sub:curvature2}

As discussed in Section~\ref{sec:LLTB}, the $\Lambda$LTB model has three arbitrary functions. We have set two of them, $m(r)$ and $t_{BB}(r)$, using a gauge choice and physical arguments. On the other hand, our particular choice of $k(r)$ is still arbitrary. Here, we study the impact, on the Hubble tension problem, of such an assumption by performing an extra analysis that uses a generalization of equation~\eqref{eq:kr}:
\begin{align} \label{eq:kr_alpha}
P_{3}(x,\alpha)=
\begin{cases}
1 &  \mbox{for } 0 \le x < \alpha \\
1 - \exp \left[\frac{1 - \alpha}{x - \alpha}\left(\frac{x - \alpha}{1 - \alpha} - 1 \right )^3 \right] & \mbox{for }  \alpha  \le x < 1\\
0 & \mbox{for } 1 \leq x 
\end{cases} , 
\end{align}
where $0 < \alpha < 1$ is a new parameter that modifies the smoothness of the transition between the inner and background regions. Sharper profiles are obtained when $\alpha$ approximates unity. Note that our main analysis with equation~\eqref{eq:kr} can be recovered by setting $\alpha=0$.

Results are shown in Table~\ref{tab:alpha}, where, for the sake of comparison, we also report the results relative to $\alpha=0$.
The addition of the parameter $\alpha$ leads to an increase in the value of $H_0^{\mathrm{L}}$ by $0.64$~km~s$^{-1}$~Mpc$^{-1}$ as compared with the previous analysis with $\alpha = 0$. This, along with the increment on the error, reduces the Hubble tension to $2.7 \sigma$. The tension on $M_B$ decrease to $3.2\sigma$.
In other words, we find a small improvement with respect to the analysis, but the $\Lambda$LTB cannot fully explain the tension.
The assumption of the generalized curvature profile of equation~\eqref{eq:kr_alpha} reduces the $\chi^2_{\mathrm{min}}$ by $0.3$ so that we obtain $\Delta \mathrm{AIC} = 1.7$ and $B_{01} = 1.9$ in favor of the simplest model with $\alpha = 0$. Namely, a weak evidence in favor of the curvature profile given by equation~\eqref{eq:kr} is found.

\subsection{Mapping the local structure of the universe} \label{sub:local} 

\begin{figure*}
\includegraphics[width=\textwidth]{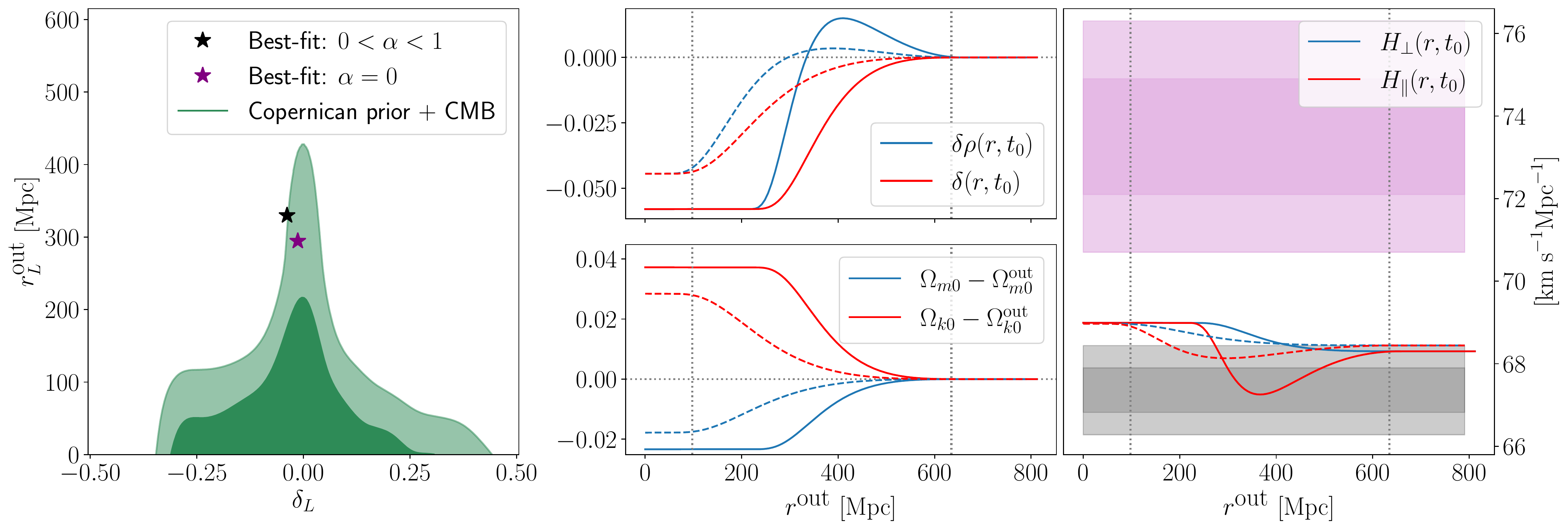}
\caption{
Characterization of our local spacetime from the best fit to all data of the $\Lambda$LTB analysis with the generalized profile with $0< \alpha<1$ (solid lines) and with $\alpha = 0$ (dashed lines).
The panel on the left shows the size $r_L^{\mathrm{out}}$ and depth $\delta_L$ of the two best-fit models as compared with the standard model expectation, which is quantified via the Copernican prior convolved with the CMB likelihood \citep[see][]{Camarena:2021mjr}.
The panels in the middle show the matter and mass density  contrasts (top) and the deviations of $\Omega_{m,0}$ and $\Omega_{k,0}$ from the $\Lambda$CDM  background (bottom) as functions of the comoving FLRW coordinate $r^{\mathrm{out}}$.
The dotted vertical lines mark the redshift range $0.023<z<0.15$ that is used to determine $H_0$.
The panel on the right shows the rates of expansion $H_{\parallel}(r,t_0)$ and $H_{\perp}(r,t_0)$ a function of $r^{\mathrm{out}}$. The purple and gray areas correspond to constraints at 68\% and 95\% confidence level of the Hubble constant as determined by the SH0ES \citep[][]{Reid:2019tiq} and Planck collaboration \citep[][]{Aghanim:2018eyx}, respectively.
See Section~\ref{sub:local} for details.\label{fig:delta_alpha}}
\end{figure*}

While  Occam's razor favors the $\Lambda$LTB model with $\alpha=0$, the generalized curvature profile is useful to map the local matter distribution.
Figure~\ref{fig:delta_alpha} shows the rates of expansion $H_{\parallel}(r,t_0)$ and $H_{\perp}(r,t_0)$ (right panel), the matter and mass density (top mid panel), and the deviations of $\Omega_{m,0}$ and $\Omega_{k,0}$ from the $\Lambda$CDM  background (bottom mid panel) as functions of the comoving FLRW coordinate $r^{\mathrm{out}}$ 
for the best fit of the analysis CMB + $M_B$ + All with equation~\eqref{eq:kr_alpha} (solid lines).
Local fluctuations in the matter density parameters were found by \citet{Colgain:2022nlb} when analyzing supernova data.
We also display the same quantities considering the best fit of our main analysis with $\alpha = 0$ (dashed lines).
The best-fit values are 
\begin{align}
\{\!\alpha,\delta_L, r^{\rm out}_L, \Omega^{\rm out}_{m,0}, H^{\rm out}_0  \!\}\!\!=\!\!\{\! 0.28, -0.038, 330, 0.304, 68.3 \! \}
\end{align}
for the case of the generalized profile of equation~\eqref{eq:kr_alpha}, and
\begin{align}
\{\delta_L, r^{\rm out}_L, \Omega^{\rm out}_{m,0}, H^{\rm out}_0    \}=\{ -0.013, 294, 0.302, 68.4  \}
\end{align}
for the case with $\alpha = 0$.

The left panel of Figure~\ref{fig:delta_alpha} shows size $r_L^{\mathrm{out}}$ and depth $\delta_L$ of the two best-fit models as compared with the standard model expectation, which is quantified via the Copernican prior convolved with the CMB likelihood \citep[see][]{Camarena:2021mjr}.
We can see that the data prefers a shallow  void with $\delta_L \approx -0.04$ and $r^{\mathrm{out}}_L \approx 300$ Mpc, which, interestingly, lies on the border of the 95\% credible region relative to the standard model expectation.

Even though the analysis including $\alpha$ allows us to map the local distribution of matter in a more general way, the local structure of the Universe could be restricted using a yet more general profile, such as an $n$-node spline function \citep{Redlich:2014gga} or a data-driven technique, possibly including anisotropic degrees of freedom.
Indeed, while our modeling is adequate to test if a local underdensity can explain away the Hubble tension, see Section~\ref{anisotropies}, it may be important to consider anisotropies when modeling a shallow structure such as the one depicted in Figure~\ref{fig:delta_alpha}.
This is also suggested by recent maps of our cosmological neighborhood \citep{Courtois:2013yfa}.
We leave this problem to the future.

\section{Conclusions} \label{sec:conclusions}

In \citet{Camarena:2021mjr}, we pursued a program to test one of the fundamental assumptions of modern cosmology: the Copernican principle.
In particular, we modeled the spacetime around us without any prior on the parameters that describe the inhomogeneity, but rather letting observations constrain the local structure. 
Our analysis showed that current cosmological data can tightly constrain radial deviations from the FLRW metric at almost the cosmic variance level. We also showed that typical constraints on the $\Lambda$CDM parameters are not weakened if one drops the Copernican hypothesis.
Here, our aim was to quantify the impact of the Copernican principle on the Hubble problem: can a non-Copernican structure explain away the Hubble tension?

In order to robustly answer this question, we put care on how to compute the Hubble constant in a inhomogeneous universe, which we parametrize via the $\Lambda$LTB model---basically a radial perturbation of a FLRW metric.
We adopted three  different definitions, which all give basically similar results. Then, in order to quantitatively conclude if the extra geometrical degrees of freedom of the $\Lambda$LTB model are favored by the data, we carried out Bayesian model selection via both the Bayes factor and the  Akaike information criterion. Finally, we considered both a flat and a curved background FLRW model.
Our results show that the $\Lambda$LTB model can successfully explain away the $H_0$ tension and is favored with respect to the $\Lambda$CDM model only if one solely considers supernovae in the redshift range that is used to fit the Hubble constant, that is, $0.023<z<0.15$.
If one considers all the supernova sample then the $H_0$ tension is not solved and the support for the $\Lambda$LTB model vanishes.
We have also carried out an analysis that adopts a more general curvature profile. We have found that the inclusion of a new parameter, that sharpen or smooth the transition between the inner inhomogeneity and the background model, does not provide a solution to the Hubble constant problem, only slightly increasing the local expansion rate.
Our results are in good agreement with previous studies and improve upon them by considering a thorougher statistical analysis and a more comprehensive set of observations.

Finally, we have used the generalized curvature profile to reconstruct our local spacetime. We have found that the best fit to current cosmological data corresponds to a shallow  void with $\delta_L \approx -0.04$ and $r^{\mathrm{out}}_L \approx 300$ Mpc, which, interestingly, lies on the border of the 95\% credible region relative to the standard model expectation.
A more generic reconstruction of the local matter distribution of the Universe could be achieved using data-driven methods. We leave the study of this topic for future research.

\section*{Acknowledgments}

It is a pleasure to thank Wessel Valkenburg for sharing \texttt{VoidDistances2020}.
DC thanks CAPES for financial support.
VM thanks CNPq and FAPES for partial financial support. CC is supported by the UK Science \& Technology Facilities Council Consolidated Grant ST/P000592/1.
This work made use of the CHE cluster, managed and funded by COSMO/CBPF/MCTI, with   financial   support   from   FINEP   and   FAPERJ, and  operating  at  the  Javier  Magnin  Computing  Center/CBPF. 
This work also made use of the Virgo Cluster at Cosmo-ufes/UFES, which is funded by FAPES and administrated by Renan Alves de Oliveira.

\section*{Author contributions}

VM and CC conceived the research question.
All authors designed the study and analysis plan.
DC led the numerical implementation of the model and observables and the MCMC exploration.
ZS contributed to the numerical implementation and MCMC exploration.
DC and VM drafted the initial version of the manuscript.
All authors critically reviewed early and final versions of the manuscript.

\section*{Data availability}

The data underlying this article will be shared on reasonable request to the corresponding author.
The \texttt{monteLLTB} code is available at \href{https://github.com/davidcato/monteLLTB}{github.com/davidcato/monteLLTB}.


\bibliographystyle{mnrasArxiv}
\bibliography{biblio}


\appendix

\section{Results with the prior on $H_0$}
\label{ap:H-M}

\begin{figure}
\centering
\includegraphics[width=\columnwidth]{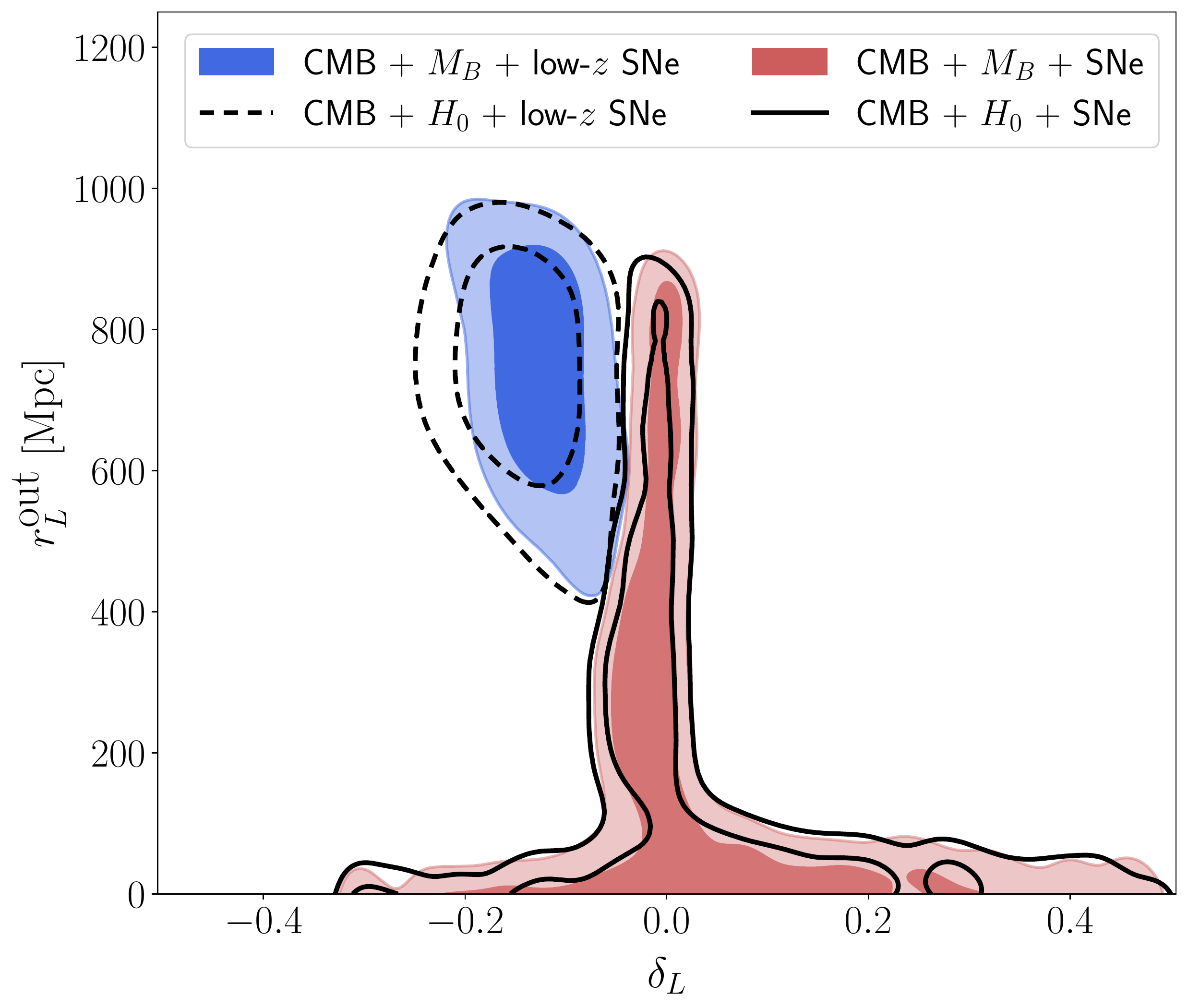}
\caption{Marginalized constraints on the effective mass density contrast $\delta_L$ and compensating scale $r_L^{\rm out}$ of the $\Lambda$LTB model in a flat background Universe at 68\% and 95\% confidence level.
\label{fig:LLTB_H}}
\end{figure}

Here, for completeness, we compare the constraints that are obtained using the prior on $M_B$ with the ones obtained using the prior on local $H_0$, see Figure~\ref{fig:LLTB_H}. We can see that the two choices provide very similar constraints thanks to the way we implemented the prediction of the local Hubble rate, see Section~\ref{sec:H0}.

\section{Qualitative interpretation of Bayes ratio and $\Delta$AIC}
\label{ap:bayes}

Tables~\ref{tab:jeffreys} and~\ref{tab:AIC} present the scales that we adopt for the interpretation of the quantitative results from model selection via the Bayes ratio and $\Delta$AIC.

\begin{table}
\centering
\begin{tabular}{>{\centering\arraybackslash} m{1.5cm} | >{\centering\arraybackslash} m{5cm}}
\hline
\hline
$\ln B_{01}$ & Strength of evidence \\
\hline
$> 5$ & Strong evidence for $\Lambda$CDM  \\ 
$[2.5, 5] $& Moderate evidence for $\Lambda$CDM  \\ 
$[1, 2.5]$ & Weak evidence for $\Lambda$CDM  \\ 
$[-1, 1] $& Inconclusive  \\ 
$[-2.5, -1]$ & Weak evidence for $\Lambda$LTB  \\ 
$[-5, -2.5]$ & Moderate evidence for $\Lambda$LTB  \\ 
$<- 5$ & Strong evidence for $\Lambda$LTB  \\ 
\hline
\end{tabular}
\caption{Conservative Jeffreys' scale~\citep{Trotta:2008qt}. \label{tab:jeffreys}}%
\end{table}

\begin{table}
\centering
\begin{tabular}{>{\centering\arraybackslash} m{1.5cm} | >{\centering\arraybackslash} m{5cm}}
\hline
\hline
$\left| \Delta \mathrm{AIC} \right| $& Level of empirical support for the model with the higher AIC \\
\hline
0 -- 2 & Substantial \\
4 -- 7 & Considerably less \\
$>10$ & Essentially none \\
\hline
\end{tabular}
\caption{Qualitative interpretation of $\Delta$AIC according to the calibrated Jeffreys' scale \citep{burnham2002practical}.
}
\label{tab:AIC}
\end{table}

\label{lastpage}
\end{document}